\documentclass[letterpaper, 10 pt, conference]{ieeeconf}  

\IEEEoverridecommandlockouts                              

\usepackage{graphics} 
\usepackage{amsmath} 
\usepackage[nocomma]{optidef} 
\usepackage[linesnumbered,lined,commentsnumbered]{algorithm2e}
\usepackage{bm}
\usepackage{tikz} 
\usetikzlibrary{math}
\usetikzlibrary{arrows.meta}
\usepackage{amssymb}  
\usepackage{graphicx}

\newcommand{\ub}[1]{\overline{ #1 }}
\newcommand{\lb}[1]{\underline{ #1 }}

\newcommand{\norm}[1]{\left\lVert #1 \right\rVert}

\newcommand{\state}{x}
\newcommand{\controls}{u}

\title{\LARGE \bf
	An Inverse Optimal Control Approach for Trajectory Prediction of Autonomous Race Cars 
}

\author{Rudolf Reiter$^{1}$, Florian Messerer$^{1}$, Markus Schratter$^{2}$, Daniel Watzenig$^{2,3}$ and Moritz Diehl$^{1,4}$
	\thanks{$^{1}$Department of Microsystems Engineering (IMTEK), University Freiburg, 79110 Freiburg, Germany
	{\tt\small \{rudolf.reiter, florian.messerer, moritz.diehl\}@imtek.uni-freiburg.de}}%
	\thanks{$^{2}$Virtual Vehicle Research Center, Inffeldgasse 21a, 8010 Graz, Austria 
	{\tt\small \{markus.schratter, daniel.watzenig\}@v2c2.at}}%
	\thanks{$^{3}$Institute of Automation and Control, Graz University of Technology, Inffeldgasse 21b, Graz, Austria. 
	}%
	\thanks{$^{4}$Department of Mathematics, University Freiburg, 79110 Freiburg, Germany
	}%
}

\begin{document}

	\maketitle
	\thispagestyle{empty}
	\pagestyle{empty}

	\begin{abstract}
		This paper proposes an optimization-based approach to predict trajectories of autonomous race cars. We assume that the observed trajectory is the result of an optimization problem that trades off path progress against acceleration and jerk smoothness, and which is restricted by constraints. The algorithm predicts a trajectory by solving a parameterized nonlinear program (NLP) which contains path progress and smoothness in cost terms. By observing the actual motion of a vehicle, the parameters of prediction are updated by means of solving an inverse optimal control problem that contains the parameters of the predicting NLP as optimization variables. The algorithm therefore learns to predict the observed vehicle trajectory in a least-squares relation to measurement data and to the presumed structure of the predicting NLP. This work contributes with an algorithm that allows for accurate and interpretable predictions with sparse data. The algorithm is implemented on embedded hardware in an autonomous real-world race car that is competing in the challenge \emph{Roborace} and analyzed with respect to recorded data.
	\end{abstract}

	\section{INTRODUCTION}
\begin{figure}
			\begin{center}
				\includegraphics[scale=0.315]{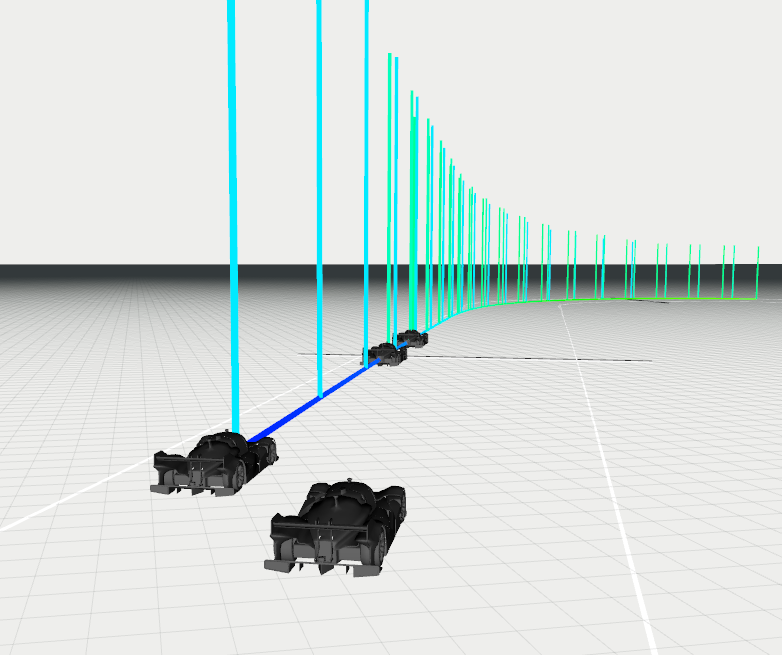}
				\caption{Presented trajectory prediction in a simulation. The height and color of the bars correspond to the predicted speed.}
				\label{fig:racecar}
			\end{center}
		\end{figure}

		In real-world autonomous driving scenarios, a core challenge is the prediction of other agents in the environment. The prediction algorithms differ related to the scenario and to the availability of data. For instance, in urban driving, a large amount of data might be available due to massive data collection of the vehicle industry. For autonomous racing tasks, there is a lack of extensive data sets, thus supervised learning of data-driven predictions is not feasible. In our research, we focus on a racing setting related to a competition called \emph{Roborace}. As part of this racing series, the participating teams develop software for the fully autonomous operation of electric race cars and are confronted with increasingly demanding challenges from one event to the other. Whereas the ego vehicle moves on a real racetrack, the state observations of the (currently) purely virtual opponent race cars are provided by the mixed-reality simulator to the car's software about 200 meters in advance. The up to six virtually present opponent cars are set up by the organizers with different racing algorithms that are supposed to race with different performance and driving style. The opponent cars are currently not considered as strategic decision makers, i.e., they are not performing game theoretic actions such as blocking. Thus, the race cars can be seen as non-interactive agents. Generally, there is no a priori knowledge available about the opponents, except of their racing intention. Therefore, it is impossible to use an a priori fully parameterized vehicle model as a basis for prediction. Furthermore, an extensive system identification is impossible due to the short time the vehicle can be observed before it needs to be overtaken. The goal of this paper is to present a method that predicts the behavior of other race cars even with sparse data. A typical scenario is shown in Fig.~\ref{fig:racecar}, where the ego race car and three other opponent cars are on a racetrack and trajectories of our presented predictor are shown with bars, corresponding to the predicted velocity.
		
		Our work starts with framing the basic and limited knowledge about the opponents as a sparsely parameterized predictor whose parameters can be estimated by a limited amount of data. Since it is known that the intention of the other opponents is time-optimal driving, the predictor is stated as a parameterized optimization problem for progress maximization, referred to as low-level nonlinear program (LLNLP) which is assumed to be solved by the other agent. The estimation of the parameters is performed by solving an inverse optimal control (IOC) problem, which enforces the optimality conditions for the LLNLP as constraints and performs least-squares optimization on the deviation of the resulting LLNLP-trajectory to the collected observed data of the particular opponent vehicle. This results in a set of parameters for the LLNLP which are locally optimal with respect to the chosen structure of the LLNLP and the observed data. In fact, the chosen formulation only finds a stationary point as opposed to an optimal point and is dependent on the initialization due to its non-convex structure, but in practice both were observed to not have a significant influence on the performance. The LLNLP is solved in real-time for each opponent, starting with an initial set of parameters, which are updated as soon as enough data is available. \\
		
		The LLNLP is used to predict the velocity along a curve, which is obtained by blending the current motion into a previously computed minimum curvature path. The parameters related to LLNLP are the constraint limits and the square penalties on the input (jerk) and the acceleration states. The estimation of the acceleration constraints is separated from the bi-level optimization problem into a separate constraint estimation QP (CQP) whose constraint estimates update both the bi-level program for the weight parameter estimation and the final LLNLP for predicting the opponent trajectories in real time. \\
		
		The performance of the described algorithm is shown with recorded data from differently driving opponent race cars in a \emph{Hardware-In-The-Loop} setting. 
		
		\subsection{Related work}
			Trajectory prediction in the domain of autonomous vehicles is dominated by data-driven approaches which are based on regression and pattern matching. This is applicable if the availability of sufficient data related to human driven vehicles on public streets is given. If interaction and sequential decision making is considered, often IOC or inverse reinforcement learning (IRL) are used. Often deep neural networks (DNNs) are used as function approximators \cite{Florin2021} and the time dependency suggests the use of recurrent neural architectures as seen in \cite{Millefiori2021, ZHANG20209, Andre2021}. Also, various other DNN architectures are used, such as convolutional neural networks \cite{nikhil2018convolutional}. If statistically qualitative data is available, these approaches work well, and even their application to real time systems as trained networks is favorable due to the high evaluation speed of DNNs. Using an optimization problem as a function approximator or even within a neural network is a field with many related research areas, ranging from reinforcement learning with an embedded MPC structure \cite{zanon2020} to generic optimization layers \cite{Kolter2019}. Using bi-level optimization to estimate the parameters of a low-level problem is used more rarely. Related to vehicle predictions, it was used in a similar approach, which focuses on urban driving scenarios and the game theoretic interaction between agents \cite{lucidgames,algames}. Furthermore, for robotic predictions \cite{Menner2021} or even human motion predictions \cite{Mombaur2010}, a bi-level problem was used. For unconstrained linear systems, \cite{Menner2018ConvexFA} the authors show that the IOC can even be stated as a convex semidefinite program. A detailed survey of vehicle prediction approaches is given in \cite{Florin2021}, although IOC appears only in the context of IRL. A general survey on bi-level optimization is given in \cite{Sinha2018}, which mentions the presented approach of solving the lower-level program by restricting it to a stationary point, especially for convex problems.
		
		\subsection{Contribution}
			 In the domain of autonomous racing, to the best knowledge of the authors, this work is the first that uses bi-level optimization together with the LLNLP for real-time trajectory prediction. Since bi-level problems are hard to solve, this work also addresses novel techniques that can be used in challenging real-world conditions such as racing. This paper follows previous work for solving motion planning problems for autonomous racing \cite{Reiter2021, Reiter2021a}.
	
	\section{Prediction architecture}
	\label{section:architecture}
\begin{figure}
			\begin{center}
				\includegraphics[scale=0.45]{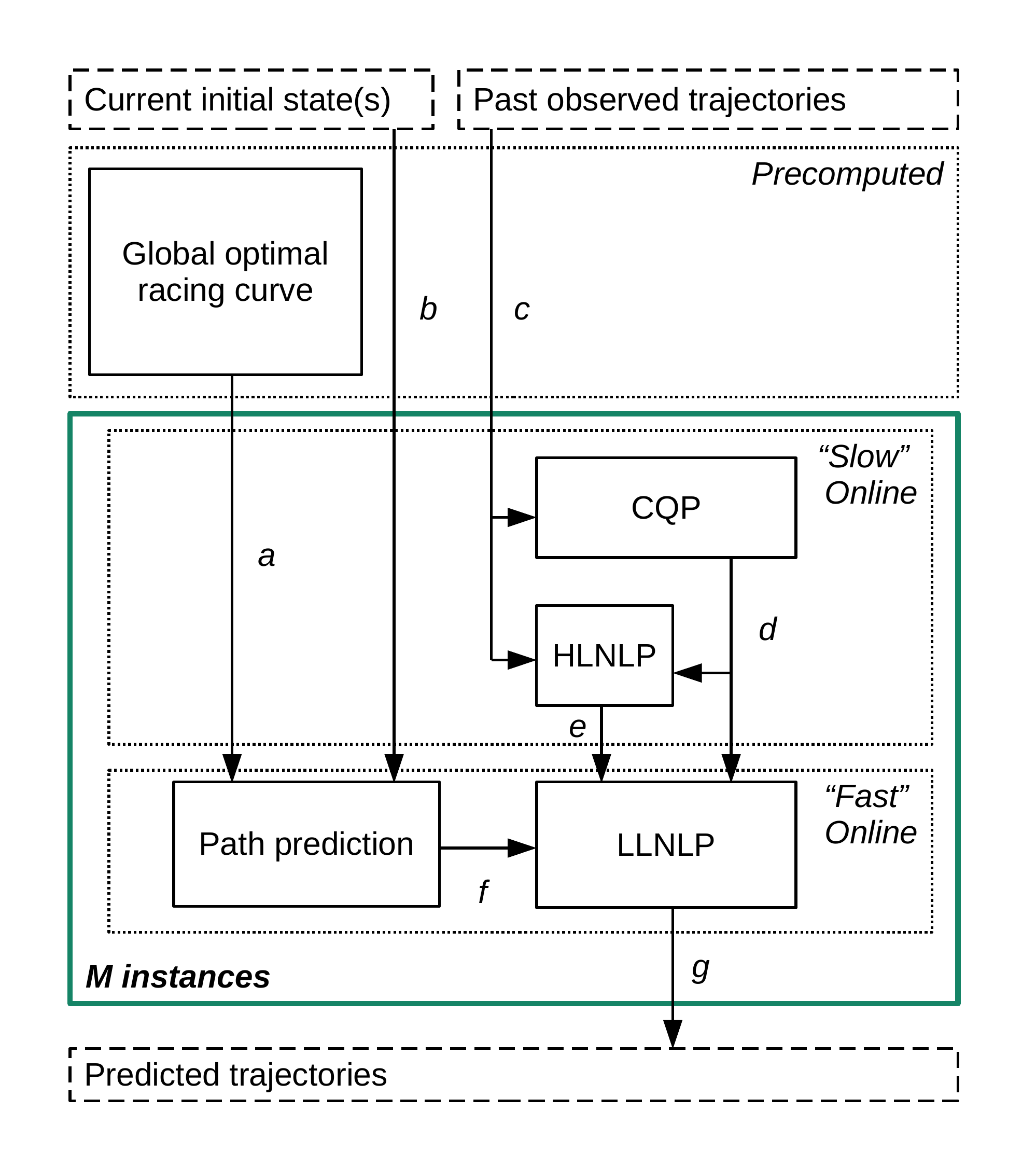}
				\caption{Algorithm architecture. \textit{(a: global racing path, b: initial state $\bar{x}_0$, c: trajectory data samples, d: constraints $a_\mathrm{max}$, e: weights $w$, f: Cartesian coordinates and curvature parameters of blended path segment $\bar{\kappa}$, g: predicted trajectory)}}
				\label{fig:architecture}
			\end{center}
		\end{figure}

		In Fig.~\ref{fig:architecture}, the architecture of the proposed algorithm is shown. The algorithm consists of an offline and an online part. The precomputations in the offline part account for the optimal racing path along the known racetrack. The online part is constructed for each opponent vehicle that is observed and is split into a slower (0.5 Hz) estimation part and a faster (10 Hz) prediction part. In the path prediction (PP) a curve is blended from the current opponent vehicle position to the precomputed racing line. The main prediction component is the LLNLP which computes the trajectory with respect to the parameterized constraints and the parameterized weights, starting at the observed current opponent value. The constraint estimator (CQP) passes its estimated constraint parameters to the high-level NLP (HLNLP) and both the CQP and the HLNLP estimate the parameters of the LLNLP. The online part is executed for each of the $M$ observed vehicles.
	
	\section{Prediction algorithm}
		In the following, the prediction algorithm is described by each component. In Sections \ref{section:path_prediction} to \ref{section:blnlp}, the main blocks of Fig.~\ref{fig:architecture} are described and in the final part the pseudocode (\ref{alg:main}) is stated.
		
		\subsection{Path prediction (PP)}
		\label{section:path_prediction}
			Given the racetrack layout, a time-optimal path $p_\mathrm{topt}(s)$ is computed by curvature minimization related to \cite{Reiter2021}. The path variable $s$ is related to the position on a reference center track line. Given the current opponent vehicle state, a linear extended constant motion path $p_\mathrm{c}(s)$ is blended into the precomputed path for $s<s_f$ with
			\begin{align}
				p_\mathrm{p}(s) = \frac{s}{s_f} p_\mathrm{topt}(s) + \frac{s-s_f}{s_f}p_\mathrm{c}(s).
			\end{align}
			For $s \geq s_f$ the prediction path is set equal to the racing path.
		\subsection{Low-level program for the trajectory prediction (LLNLP)}
		\label{section:llnlp}
			The path predictor predicts the curve that is described by its path length $s$ and the associated curvature $\kappa(s)=\frac{d\phi}{ds}$. The curvature is described by a piece-wise \emph{linear} polynomial and parameterized to interpolate $N_\kappa$ precomputed values $\kappa_i$ for the curvature along the path segment. The values are summarized as $\bar{\kappa}=\begin{bmatrix}\kappa_i&\ldots&\kappa_{N_\kappa-1}\end{bmatrix}$. Details on the computation can be found in \cite{Reiter2021}. Note that the \emph{linear} interpolation leads to discontinuous derivatives in the inequality constraints and violates the condition of twice continuously differentiable functions required for second order NLP algorithms. Nevertheless, we empirically found a speedup of a factor of 100 to 1000 compared to \emph{bsplines} as interpolating polynomials, with practically no convergence problems.
						
			The LLNLP predicts the estimated velocity along this curve by solving an optimal control problem which consists of a linear discrete model $F(x_k,u_k,\Delta t)$, acceleration constraints $h_a(\state_k, \bar{\kappa}, a_\mathrm{max})$ and state constraints $\lb{\state}$ and $\ub{\state}$. Since the path is given, the predicted motion along the curve is described by means of three chained integrators, where the input $u$ is the jerk. The state vector consequently consists of the path progress $s$, the velocity $v$ and the acceleration $a$ with $x=\begin{bmatrix}s&v&a\end{bmatrix}^\top \in \mathbb{R}^3$. Since the integrator chain is a linear system, the discretization (zero-order-hold controls) of the dynamics can be computed exactly by matrix exponentials and leads to the affine function $F(x_k,u_k,\Delta t) = A(\Delta t)x_k+B(\Delta t)u_k$. The only constraint captured in the box constraints $\lb{\state}\preccurlyeq \state_k \preccurlyeq\ub{\state}$ is the limitation of the speed $v$ to $v_\mathrm{max}$ and to positive values.
			The optimal control problem is discretized in $N-1$ intervals using discrete multiple shooting and solved by sequential quadratic programming using the real-time NMPC solver \emph{acados} \cite{Verschueren2019}. To account for the progress maximizing requirement for the resulting trajectory, a linear negative cost $q_n=\begin{bmatrix}-1&0&0\end{bmatrix}^\top$ for the last discrete position is used. The matrix $W=\mathrm{diag}(\begin{bmatrix}0&0&w_\mathrm{acc}\end{bmatrix})$ and the scalar $R=w_\mathrm{jerk}$ are the weights that describe the motion of the predicted opponent vehicle in the presented structure, if no constraints are active. Finding the values of $w_\mathrm{acc}$ and $w_\mathrm{jerk}$ is the objective of the HLNLP component. Slack variables $s_\mathrm{LL}=\begin{bmatrix}s_{\mathrm{LL},0},&\ldots,&s_{\mathrm{LL},N}\end{bmatrix}\in\mathbb{R}^{8\times N}$ with weights $\alpha_1,\alpha_2$ are added for the online forward implementation to account for the robustness of the SQP algorithm. We can then state the lower-level problem $P_\mathrm{LL}(w, \bar{\state}_0,\bar{\kappa},a_\mathrm{max})$ as
			
\begin{figure}
				\begin{center}
					\includegraphics[scale=0.8]{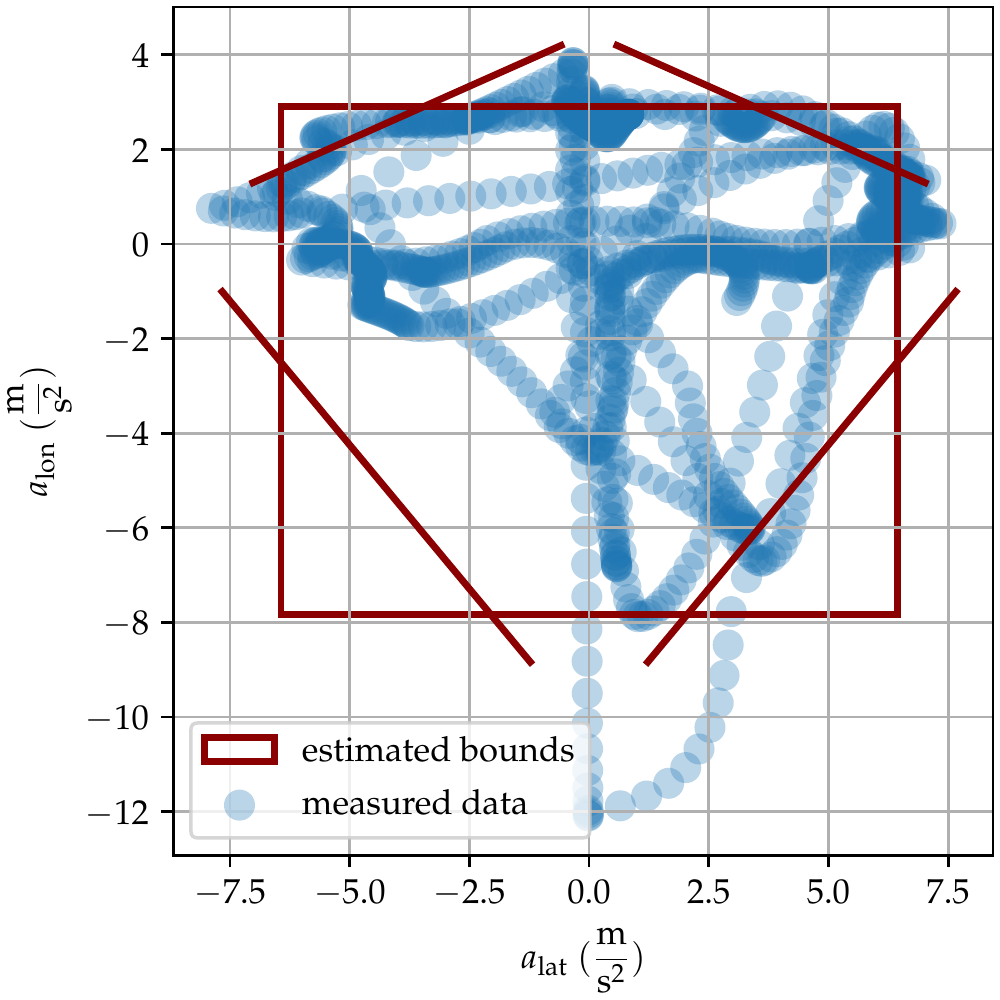}
					\caption{Acceleration constraint estimation. In total 8 constraints are fitted as a convex polytope to measurement data.}
					\label{fig:constraints_estimation}
				\end{center}
			\end{figure}

			{\begin{small}\begin{mini}
				{\begin{subarray}{c}
						\state_0, \ldots, \state_N,\\
						\controls_0, \ldots, \controls_{N-1}\\
						s_0, \ldots , s_N
				\end{subarray}}			
				{\sum_{k=0}^{N-1} \norm{\state_k}_{2,W}^2 + \norm{\controls_k}_{2,R}^2 + q_N^\top\state_{N}  }		
				{\label{eq:llnlp}} 
				{} 
				\breakObjective{ + \sum_{k=0}^{N} \alpha_1 \mathbf{1}^\top s_{\mathrm{LL},k}+ \alpha_2 \norm{s_{\mathrm{LL},k}}_2^2 }
				\addConstraint{\state_0}{= \bar{\state}_0}{},
				\addConstraint{\state_{k+1}}{= F(\state_k, \controls_k, \Delta t),}{ k=0,\ldots,N-1,}
				\addConstraint{\hspace{-0.5cm}\lb{\state}}{\preccurlyeq \state_k \preccurlyeq \ub{\state}}{},
				\addConstraint{0}{\preccurlyeq h_a(\state_k,  \bar{\kappa}, a_\mathrm{max})+s_{\mathrm{LL},k}}{},
				\addConstraint{0}{\preccurlyeq s_{\mathrm{LL},k},}{\quad k=0,\ldots,N,}
			\end{mini}\end{small}} 
			where $\mathbf{1}$ is a vector of all $1$'s of appropriate size. The acceleration constraints $h_a(\state_k, \bar{\kappa}, a_\mathrm{max})$ approximate the friction, throttle and breaking boundaries of the vehicle by means of a polytope in the space of the two-dimensional acceleration vector $a(x_k, \bar{\kappa})=\begin{bmatrix}a_\mathrm{lat}(x_k, \bar{\kappa})&a_\mathrm{lon}(x_k)\end{bmatrix}$ which are often related to the "Kamm's circle". The polytope is typically symmetric to the longitudinal axis. It is chosen such that it consists of box-constraints along the axes and diagonal constraints that are parallel to the lines described by the connection of the axis aligned maximum values. Consequently, the diagonal constraints depend on values of the axis aligned constraints. The presented approach computes the axis aligned constraints first and uses those values as inputs to the diagonal constraints. An example of the fitted acceleration constraints can be seen in Fig.~\ref{fig:constraints_estimation}. 
			Therefore, 8 linear constraints arise, where 6 of them are pair-wise symmetric. The only non convexity in (\ref{eq:llnlp}) emerges from the dependency of $a_\mathrm{lat}(x_k)=- v_k^2\kappa(s_k, \bar{\kappa})$. The acceleration constraints $h_a(\state_k, \bar{\kappa}, a_\mathrm{max})$ can be stated as 
			\begin{subequations}	
			\begin{align}
			\label{eq:llnlp_constraints}
			h_a(\state_k,  \bar{\kappa}, a_\mathrm{max}) 			&= a_{\mathrm{max}} - \mathrm{diag}(d_{\mathrm{len}}) D a(x_k, \bar{\kappa})  \\
			\bar{a} 			&= \sqrt{a_\mathrm{lat,max}^2 + a_\mathrm{lon,max}^2}\\
			d_{\mathrm{len}}^\top	&= \begin{bmatrix}1&1&1&1&\bar{a}&\bar{a}&\bar{a}&\bar{a}\end{bmatrix}
			\end{align}
			\begin{align}
			D					&= \begin{bmatrix}  1 & 0 \\ 
													-1 & 0 \\ 
													0 & -1 \\ 
													0 & -1 \\ 
													a_\mathrm{lon,max} & a_\mathrm{lat,max} \\ 
													-a_\mathrm{lon,max} & a_\mathrm{lat,max} \\ 
													a_\mathrm{lon,min} & -a_\mathrm{lat,max} \\ 
													-a_\mathrm{lon,min} & -a_\mathrm{lat,max} \\\end{bmatrix},\text{ }
			a_{\mathrm{max}} = \begin{bmatrix}  a_{\mathrm{lat,max}}\\
										a_{\mathrm{lat,max}}\\
									    a_{\mathrm{lon,max}}\\
									    a_{\mathrm{lon,min}}\\
										a_{\mathrm{q,north}}\\
										a_{\mathrm{q,north}}\\
										a_{\mathrm{q,south}}\\
										a_{\mathrm{q,south}}\end{bmatrix}
			\label{eq:constraints}
			\end{align}
			\end{subequations}
			The term $\mathrm{diag}(d_{\mathrm{len}}) D$ collects the row vectors with unit length that represent direction vectors that are used to project the acceleration vector and to constrain to the consecutive projected scalar value. Obviously, the second part of the vector $d_{\mathrm{len}}$ and the lower part of the matrix $D$ shows the dependency on the axis aligned constraints.
			The maximum acceleration values are collected in the vector $a_{\mathrm{max}}$, the initial observed state $\bar{x}_0$, together with the weights $w^\top=\begin{bmatrix}w_{\mathrm{jerk}} & w_\mathrm{acc}\end{bmatrix}$ are the input parameters of the LLNLP. The parameters that are not subject to be changed by the estimation within the HLNLP are summarized as $p^\top = \begin{bmatrix}\bar{x}_0 & a_{\mathrm{max}}^\top & \bar{\kappa}\end{bmatrix}$.
		
		\subsection{Quadratic program for constraint estimation (CQP)}
		\label{section:cqp}
			In order to remove computational complexity from the HLNLP, the constraint estimation was separated. Even with fixed constraints, the structure of the HLNLP is highly non-convex and challenging to solve. 
			By means of a Kalman-filter-based vehicle state estimation, the observed accelerations $a_i$ are computed and stored as a data set of $N_a$ data samples in $\mathbb{R}^2$. Those acceleration data are used to fit linear constraints. The projection $e_k^\top D a_i$ for the estimation of the linear constraint $c_k$ is performed for the 8 constraints. The vector $e_k$ represents the k-th unit vector in $\mathbb{R}^8$. Since the measured data is noisy, it requires a robust estimation of the constraints which accounts for outliers. This is achieved by the quadratic program (\ref{eq:cqp_1}), where the estimated constraint violation is penalized linearly and realized by means of a hinge loss $h(x)=\text{max}(0, x)$. This function adds no costs if the measured value is lower than the constraint, and penalizes linearly otherwise. The deflection of the constraint is penalized quadratically with a weight $\omega$ and a minimum at the prior estimated value $\hat{c}_{k}$. Since the prior estimated value acts as a lower bound and would not decrease during iterations, the value is lowered by a factor $r<1$ in each iteration for the axis aligned constraints with $\hat{c}_{k}\gets r \hat{c}_{k}$. For the diagonal constraints, $\hat{c}_{k}$ is chosen as the distance of the diagonal line of the origin. The problem formulation for estimating constraint $k$ in the bounding polytope with 8 linear constraints is stated as 
			{\begin{mini}
				{c_k \in \mathbb{R}}			
				{\frac{1}{N_a}\sum_{i=0}^{N_a-1} \text{max}(0, e_k^\top D a_i - c_k) } 
				{\label{eq:cqp_1}} {}
				\breakObjective{+ \omega(c_k-\hat{c}_{k})^2}.
			\end{mini}}
			The problem can be formulated into a smooth quadratic program using slack variables $\zeta$ for implementing the hinge function which leads to the formulation
			{\begin{mini}
				{\begin{subarray}{c}
						c_k \in \mathbb{R},\\
						\zeta \in \mathbb{R}^{N_a} 
				\end{subarray}}			
				{\omega(c_k-\hat{c}_{k})^2 + \frac{1}{N_a}\sum_{i=0}^{N_a-1} \zeta_i} 
				{\label{eq:cqp_2}} {}
				\addConstraint{0}{\leq \zeta_i}{}
				\addConstraint{e_k^\top D a_i-c_k}{\leq \zeta_i}{\quad i=0,\ldots,N_a-1}.
			\end{mini}}
			The solutions of the CQP are directly used as acceleration constraints $a_\mathrm{max}$ in (\ref{eq:constraints}).

		\subsection{Bi-level program for the LLNLP parameter estimation (HLNLP)}
		\label{section:blnlp}
			The HLNLP fits the LLNLP with the parameters derived from the CQP to observed measurement data $\bar{x}$ with a least-squares error measure. The observed trajectory might differ at the last points from the predicted trajectory, even if the true parameters were used, since the controller of the observed vehicle most likely had adapted to even further distant constraints like a sharp curve. To account for this structural uncertainty, the weight matrix $Q_k$ is linearly reduced to zero for the final $N_R$ points. Problem \eqref{eq:hlnlp_1} shows the basic structure of the problem. The optimization variables are the estimated trajectory $x$ of the LLNLP and the weighting parameters $w$. To account for the iterative estimation of the parameter $w$, the previously estimated parameter $\hat{w}$ together with the associated weight matrix $P$ is used as an arrival cost, as shown with MHE in \cite{Rawlings2017}. To simplify the algorithm, the weight matrix is set constant.
			The basic structure of the problem can be written as
			{\begin{mini}
				{\state,\controls,w}			
				{\sum_{k=1}^{N_T-1} \norm{\state_k - \bar{\state}_k}_{2,Q_k}^2 + \norm{w - \hat{w}}_{2,P^{-1}}^2}
				{\label{eq:hlnlp_1}} 
				{} 
				\addConstraint{x, u}{\in \mathrm{argmin }P_\mathrm{LL}(w, \bar{\state}_0,\bar{\kappa},a_\mathrm{max})}{},
				\addConstraint{w}{\succcurlyeq 0}{},
			\end{mini}}
			where $\state \in \mathbb{R}^{N_x \times N_T},
			\controls \in \mathbb{R}^{N_u \times (N_T-1)}$ and $w \in \mathbb{R}^2$. The optimization variables are the estimated trajectory $x$ of the LLNLP and the weighting parameters $w$.		
		
			The low-level program $P_\mathrm{LL}(w, \bar{\state}_0,\bar{\kappa},a_\mathrm{max})$ in (\ref{eq:llnlp}) can be written as  
			{\begin{mini}
				{z \in \mathbb{R}^{N_z}}			
				{f_\mathrm{LL}(z, w)}
				{\label{eq:llnlp_short}} 
				{} 
				\addConstraint{g_\mathrm{LL}(z)}{=0}{},
				\addConstraint{h_\mathrm{LL}(z, \bar{\state}_0,\bar{\kappa},a_\mathrm{max})}{\succcurlyeq 0}{},
			\end{mini}}
			with $z=\begin{bmatrix}\mathrm{vec}(x)^\top&\mathrm{vec}(u)^\top\end{bmatrix}^\top$ and $N_z = N_x  N_T + N_u  (N_T - 1)$.
			The domains and co-domains of the functions are $f_\mathrm{LL}:\mathbb{R}^{N_z \times N_w} \rightarrow \mathbb{R}$, $g_\mathrm{LL}:\mathbb{R}^{N_z} \rightarrow \mathbb{R}^{N_T N_x}$ and $h_\mathrm{LL}:\mathbb{R}^{N_z} \rightarrow \mathbb{R}^{N_T N_{h,\mathrm{LL}}}$, where $N_{h,\mathrm{LL}}=10$ in this case, with two bounds on the velocity state and 8 acceleration constraints. The number of weights corresponding to the smoothness is $N_{w}=2$. The constraints $a_\mathrm{max}$ are parameters and updated by means of the estimation of the CQP.\\
			To solve the problem, the bi-level problem can be formulated as an NLP which is summarized as
			
			{\begin{small}\begin{mini!}
					{\state,\controls,w,\tau,\lambda,\mu,s}			
					{\sum_{k=0}^{N_T-1} \norm{\state_k - \bar{\state}_k}_{2,Q_k}^2 + q_\tau \tau}
					{\label{eq:hlnlp_2}} 
					{} 
					\breakObjective{ +  \beta_1 \mathbf{1}^\top s+ \beta_2 \norm{s}_2^2 + \norm{w - \hat{w}}_{2,P^{-1}}^2}
					\addConstraint{0=}{\nabla_z f(z,w) - \nabla_z  g_\mathrm{LL}(z) \lambda}{}\nonumber
					\addConstraint{}{-\nabla_z  h_\mathrm{LL}(z,p) \mu,}{}
					\label{eq:hlnlp_2_kkt_diff}
					\addConstraint{0\preccurlyeq}{w,}{}
					\addConstraint{0=}{g_\mathrm{LL}(z),}{}
					\label{eq:hlnlp_2_kkt_eq}
					\addConstraint{0 \leq}{ \tau,}{}
					\label{eq:hlnlp_2_kkt_ieq1}
					\addConstraint{0\preccurlyeq }{ \mu,}{}
					\label{eq:hlnlp_2_kkt_ieq2}
					\addConstraint{0\preccurlyeq}{ h_\mathrm{LL}(z, p)+s,}{}
					\label{eq:hlnlp_2_kkt_ieq21}
					\addConstraint{\tau \geq }{\mu_i h_{\mathrm{LL},i}(z,p), \quad i=0,\ldots,N_{h,\mathrm{LL}}-1 ,}{},
					\label{eq:hlnlp_2_kkt_ieq3}
					\addConstraint{s\succcurlyeq}{ 0,}{}
			\end{mini!}\end{small}} where 
			$\state \in \mathbb{R}^{N_x \times N_T},
			\controls \in \mathbb{R}^{N_u \times (N_T-1)},
			w \in \mathbb{R}^2,
			s \in \mathbb{R}^{N_T N_{h,\mathrm{LL}}}, 
			\tau \in \mathbb{R},
			\lambda \in \mathbb{R}^{N_T N_x},$ and $
			\mu \in \mathbb{R}^{N_T N_{h,\mathrm{LL}}}$.
			
			We enforce a stationary point in the LLNLP as constraint in the high-level problem by enforcing the KKT conditions by means of constraints which are stated in (\ref{eq:hlnlp_2_kkt_diff}-\ref{eq:hlnlp_2_kkt_ieq3}). For this aim, additional optimization variables arise that are the dual variables $\lambda$ and $\mu$. A major challenge here is to account for the highly non-convex complementarity conditions arising from the inequalities of the LLNLP. Therefore, a relaxed problem is stated within the constraints, which is lower bounded by the actual complementarity condition and upper bounded by its relaxed version related to the interior point approach as seen in (\ref{eq:hlnlp_2_kkt_ieq1}-\ref{eq:hlnlp_2_kkt_ieq3}). If the complimentarity is relaxed too much, the estimation of the weight parameters can become wrong. Consequently, the relaxing parameter $\tau \in \mathbb{R}$ is also integrated as an optimization variable into the HLNLP and initialized with a "high" value (e.g. 1.0). A high value for $q_\tau$ together with the linear penalty of $\tau$ is used to achieve the exact complementarity constraints. Slack variables $s$ account for infeasibilities.			
			
			The number of primal variables in the high-level program, which are $N_\mathrm{var,HL}=2N_xN_T+N_u(N_x-1)+2N_hN_T$ rises notably compared to the low-level program, which are $N_\mathrm{var,LL}=N_xN_T+N_u(N_x-1)$, but is of the same complexity w.r.t. $N_x, N_T$ and $ N_h$. 
			This program is solved using the interior point solver \emph{IPOPT} \cite{Waechter2006} formulated in CasADi \cite{Andersson2019}, which again uses a relaxation of the problem in order to account for the inequality constraints. By using the presented formulation, we can explicitly account for the accuracy of the complementarity constraint in the stationary point of the low-level program. Note that the Hessian of the HLNLP actually contains third-order derivatives of the original LLNLP, thus posing the condition of three times differentiable smooth functions in the LLNLP.\\
			
		\subsection{Algorithm}
		\label{section:algorithm}
			Algorithm (\ref{alg:main}) describes the sequential interaction of the components with respect to the architecture in Fig.~\ref{fig:architecture}. The solvers \emph{CQP} and \emph{HLNLP} are executed as threads that update the estimation values in a lower frequency than the main predicting solver \emph{LLNLP}, together with the path prediction \emph{PP}.
		
			\begin{algorithm}
				\label{alg:main}
				\SetKwData{Left}{left}\SetKwData{This}{this}\SetKwData{Up}{up}
				\SetKwFunction{Union}{Union}\SetKwFunction{FindCompress}{FindCompress}
				\SetKwInOut{Input}{input}\SetKwInOut{Output}{output}
				\Input{Initial weights and constraints $\hat{w}$, $c_k$,\\
					Observed state measurements $\bar{x}_0$}
				\Output{Predicted trajectory $x_\mathrm{pred}$}
				{HLNLPsolved$\gets$True}\;
				{CQPsolved$\gets$True}\;
				\While{True}{
					\If{CQPsolved}
					{
						CQPsolved$\gets$False\;
						$\bar{x}\gets$last $N_a$ state samples\;
						$\bar{\kappa} \gets $\texttt{curv}$(\bar{x})$\;
						$\hat{c}_k\gets r c_k \quad k=0,\ldots3$\;
						$\hat{c}_k\gets \texttt{dist}(\hat{c}) \quad k=4,\ldots7$\;
						Set CQP parameters $\hat{c_k}$, $\omega$, $a(\bar{x},\bar{\kappa})$\; 
						Start CQP solver (Updates: CQPsolved, $c_k$)\;
						
					}
					\If{HLNLPsolved}
					{
						HLNLPsolved$\gets$False\;
						$\bar{x}\gets$ last $N_T$ state samples\;
						$\bar{\kappa} \gets \texttt{curv}(\bar{x})$\;
						{$\hat{w}\gets w$}\;
						$a_{\mathrm{lat},k}\gets c_{k} \quad k=0,\ldots7$\;
						Set HLNLP parameters $a_{\mathrm{lat}}$, $\bar{x}$, $\bar{\kappa}$, $\hat{w}$\; 
						Start HLNLP solver (Updates: HLNLPsolved, $w$)\;
						
					}
					$\bar{x}_0 \gets $ State measurement input\;
					$a_{\mathrm{lat},k}\gets c_{k} \quad k=0,\ldots7 $\;
					
					$\bar{\kappa} \gets PP(\bar{x}_0)$\;
					$x_\mathrm{pred} \gets $solve LLNLP$(\bar{x}_0, \bar{\kappa}, w, a_{\mathrm{lat}})$\;
					
				}
				\caption{IOC Prediction}
			\end{algorithm}
	
	\section{Results}
		The algorithm was tested with recorded data. Qualitatively, these tests fully describe the performance of the algorithm. Nevertheless, the embedded performance, especially the real-time performance of the LLNLP was proven in several real racing events. This shows that the proposed algorithm can work in embedded real-world systems.
	
		\subsection{Hardware and Software Setup}
			The proposed LLNLP was tested on a race car hardware (Section \ref{section:results_online}) including the NVIDIA DrivePX 2 in a Docker environment with Ubuntu 20.04. This electronic control unit (ECU) provides two CPUs (4x ARM Denver, 8x ARM Cortex A57) and two GPUs (2x Tegra X2, 2x Pascal GPU). The open-source \emph{OSQP} solver \cite{osqp} was used in a mixed Python/C++ ROS-framework for solving problem (\ref{section:cqp}) using CasADi \cite{Andersson2019} as an interface. CasADi was also used as an interface together with \emph{IPOPT} \cite{Waechter2006} to solve the HLNLP of Section \ref{section:blnlp}. The time critical real-time estimation related to the LLNLP (Section \ref{section:llnlp}) was performed using acados \cite{Verschueren2019} as an NLP solver. For each opponent car, a separate solver was created which was executed as a thread, updating a data-structure that contained the most recent prediction. The full algorithm was tested offline (Section \ref{section:results_offline}) in simulations with an Alienware m-15 Notebook and an Intel Core i7-8550 CPU (1.8 GHz).The parameters used for the evaluation are shown in Table~\ref{table:parameters}. 
			\begin{table}
				\begin{center}
					\caption{Parameter Settings}
					\begin{tabular}{|c | c || c | c |} 
						\hline
						Parameter & Value & Parameter & Value \\  
						\hline\hline
						
						$c_{k,0}\ldots c_{k,3}$ & $5,$ $5,$ $2.5,$ $ \text{-}5 $  m/s & $s_f$ & $300$m \\ 
						\hline
						
						$w_0^T$ &  [0.5  0.2] &  $N$ & 111\\ 
						\hline
						
						$\omega$ & 12.5 &  $\Delta T$ (LL) & $0.1$s\\ 
						\hline
						
						$N_a$ & $10^3$ & $\Delta T$ (HL) & $1$s \\ 
						\hline
						
						$N_\kappa$ & $700$ & $\alpha_1,$ $ \alpha_2$ & $10^4 ,$ $10^8$ \\ 
						\hline	
						
						$N_T$ & $25$s & $\beta_1,$ $ \beta_2$ & $ 10^5 ,$ $10^6$ \\ 
						\hline
						$P$ & diag([$ 2\cdot 10^{-7}$  $ 9 \cdot10^{7}$]) &$q_\tau$ & $10^{7}$\\ 
						\hline	
					\end{tabular}
					\label{table:parameters}
				\end{center}
			\end{table}
			In Table~\ref{table:solver_statistics}, the time statistics of the different optimization parts are shown and in Table~\ref{table:solver_settings} the relevant settings are given. Notably, the LLNLP was failing in 2 out of 1000 randomly parameterized test runs, which was due to the linear interpolation of the curvature as described in Section \ref{section:llnlp}. This failure rate is out-weighted in practice by the enormous speed gain of a linear interpolation. 
			\begin{table}
				\begin{center}
					\caption{Component settings}
					\begin{tabular}{|c || c | c | c |} 
						\hline
						Component & Samples/Nodes & Notes & Runs \\  
						\hline\hline
						PP & 150 &  & 1e3 \\ 
						\hline
						CQP & 500 &  Eval. per constraint (1/5) & 1e2\\ 
						\hline
						HLNLP & 35 &  $\Delta t$=1s & 50\\ 
						\hline
						LLNLP & 60 & $\Delta t$=0.1s & 1e3 \\ 
						\hline		
					\end{tabular}
					\label{table:solver_settings}
				\end{center}
			\end{table}
			\begin{table}
				\begin{center}
					\caption{Solver timing statistics}
					\begin{tabular}{|c || c | c | c |c |} 
						\hline
						Component & Solver &$t_{max}$ (ms)& $t_{ave}$ (ms)& fail rate (\%) \\  
						\hline\hline
						PP & none &  $<1$ & $<1$ & 0 \\ 
						\hline
						CQP & OSQP &  15.5 & 8.1 & 0 \\ 
						\hline
						HLNLP & IPOPT &  6237 & 520 & 5 \\ 
						\hline
						LLNLP & \begin{tabular}[x]{@{}c@{}}acados\\hpipm(QP)\end{tabular} &  2748 & 91 & 0.2 \\ 
						\hline		
					\end{tabular}
					\label{table:solver_statistics}
				\end{center}
			\end{table}

		\subsection{Performance analysis with recorded data}
		\label{section:results_offline}
			\subsubsection{Validation of the CQP}
			\label{section:results_offline_cqp}
				Fig.~\ref{fig:constraints_estimation} shows the estimation of constraints related to 1000 recorded acceleration data samples. The acceleration was computed out of the observed and estimated trajectory state.
				Obviously, the constraints of any observed race car acceleration data could be of any shape, but the representational capacity of the constraint assumptions have to be traded off for a fast and reliable real-time execution in the LLNLP. According to our experience, the approximation with either 4 (box only) or 8 (adding diagonals) constraints achieved the best performance.
			\subsubsection{Validation of the velocity profile estimation}
\begin{figure}
					\begin{center}
						\includegraphics[scale=0.85]{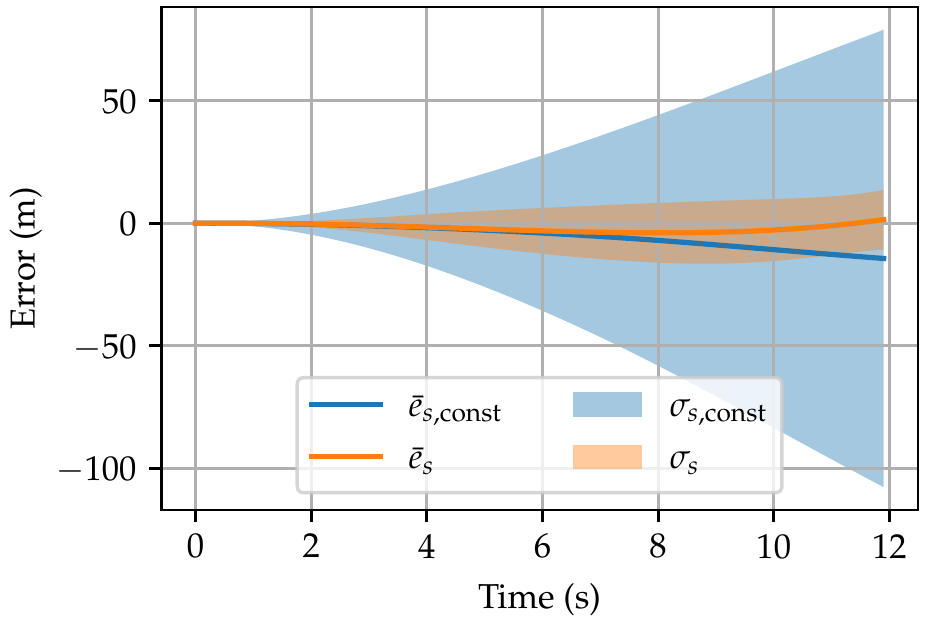}
						\caption{Mean and standard deviation of the position estimation errors for the constant velocity estimator $\bar{e}_{s,\mathrm{const}} / \sigma_{s,\mathrm{const}} $ and the presented algorithm $\bar{e}_s / \sigma_s$ along the path obtained from the PP component evaluated on recorded data. }
						\label{fig:estimation_stats_position}
					\end{center}
				\end{figure}

\begin{figure}
					\begin{center}
						\includegraphics[scale=0.85]{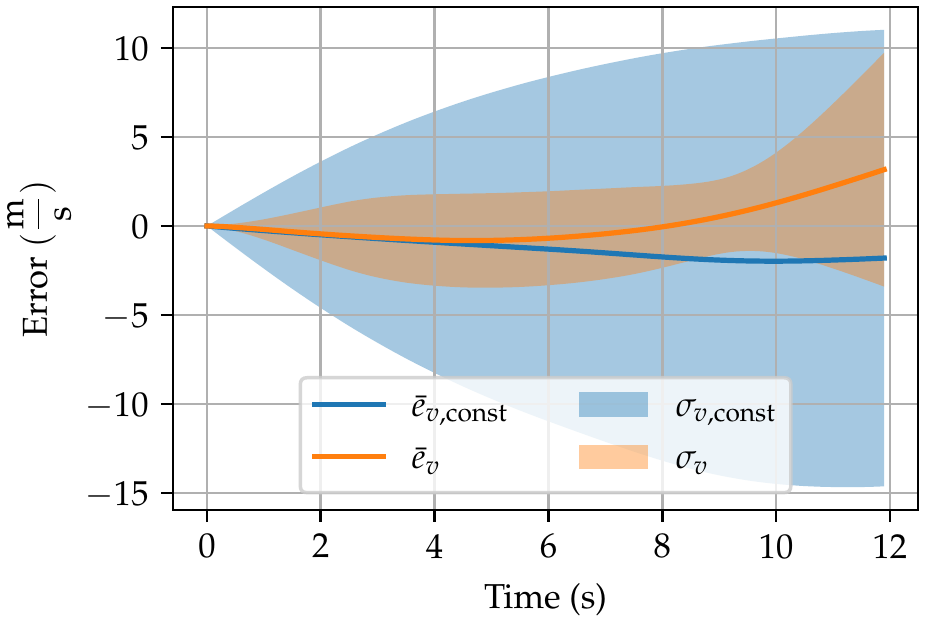}
						\caption{Mean and standard deviation of the velocity estimation errors for the constant velocity estimator $\bar{e}_{v,\mathrm{const}} / \sigma_{v,\mathrm{const}} $ and the presented algorithm $\bar{e}_v / \sigma_v$ along the path obtained from the PP component evaluated on recorded data. }
						\label{fig:estimation_stats_velocity}
					\end{center}
				\end{figure}

				Using the same recorded real-world trajectory as in \ref{section:results_offline_cqp} and also its estimated constraints $a_\mathrm{max}$ as seen in Fig.~\ref{fig:constraints_estimation}, we use the HLNLP to estimate the parameters $w$. All estimated parameters together are then forwarded to the LLNLP, which predicts the velocity and the progress along the given curve by solving the nonlinear program. The results are compared to the standard constant velocity predictor, which is often used in robotic applications \cite{Schoeller2019} and that assumes a vehicle progression with the measured constant velocity. In Fig.~\ref{fig:estimation_stats_position}, the mean position error of the presented algorithm $\bar{e}_s$ is compared to the mean position error of the constant velocity predictor $\bar{e}_{s,\mathrm{const}}$. Furthermore, the standard deviations $\sigma_s$ and $\sigma_{s,\mathrm{const}}$ are compared respectively.
				In Fig.~\ref{fig:estimation_stats_velocity} the prediction velocity error $\bar{e}_v$ and its standard deviation $\sigma_{v}$ of the presented algorithm are compared to the mean error and standard deviation of the velocity of the constant velocity estimator, that are $\bar{e}_{v,\mathrm{const}}$ and $\sigma_{v,\mathrm{const}}$
				The presented algorithm outperforms the constant velocity predictor significantly, although in a short prediction horizon the errors are similar.

		\subsection{Validation of the full algorithm}
		\label{section:results_online}
\begin{figure}
			\begin{center}
				\includegraphics[scale=0.8]{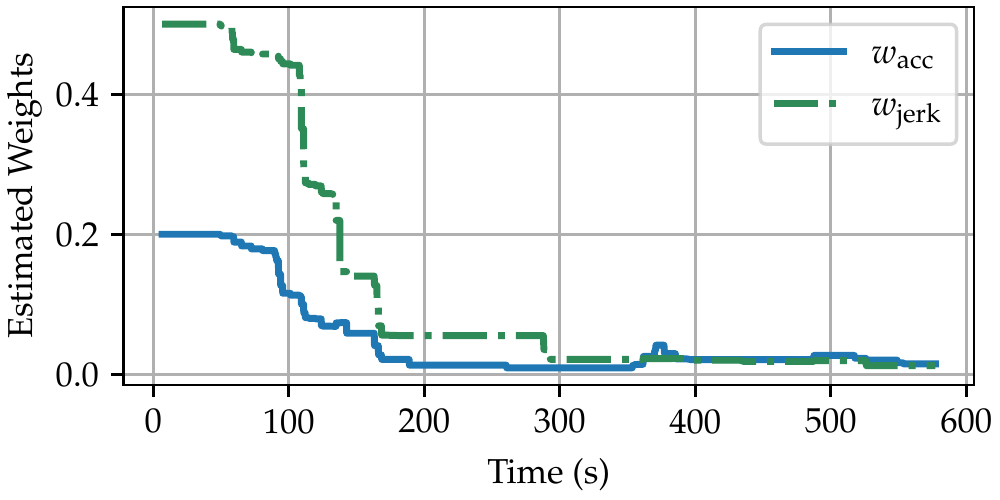}
				\caption{Weight estimates for the jerk and acceleration weighting obtained by the HLNLP component. Low weights correspond to aggressive driving that is only limited by the velocity and acceleration constraints.}
				\label{fig:full_estimated_weights}
			\end{center}
		\end{figure}

\begin{figure}
			\begin{center}
				\includegraphics[scale=0.8]{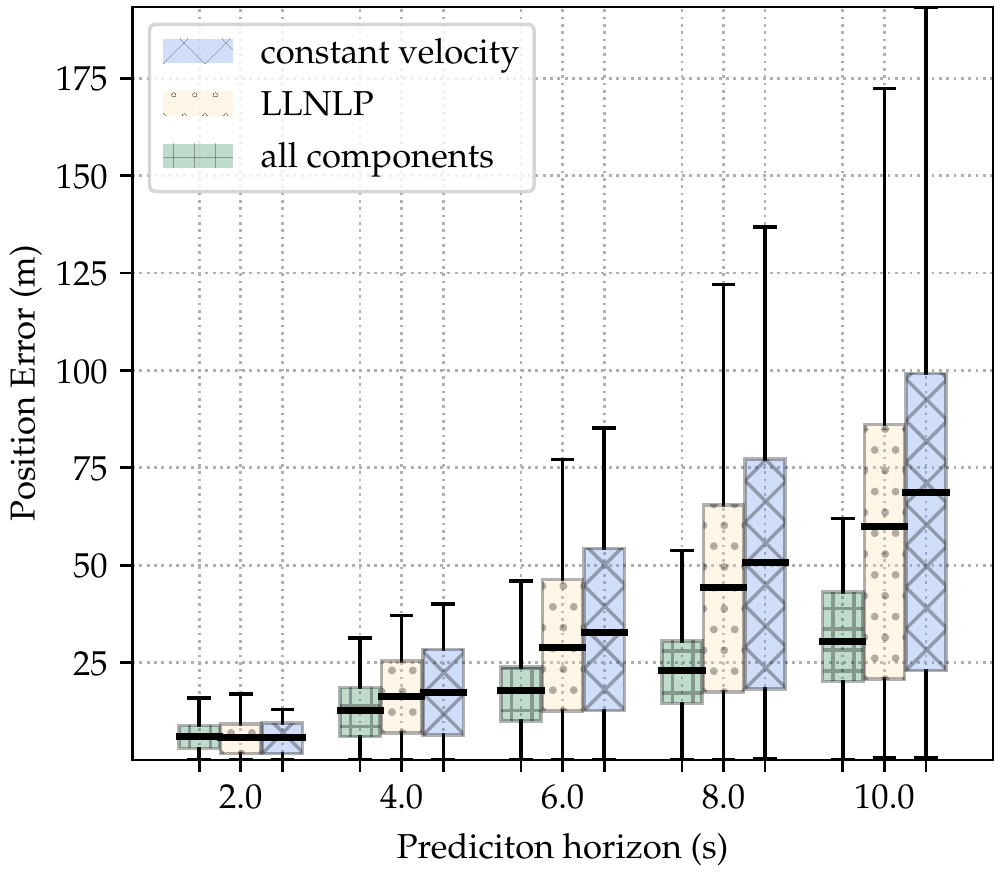}
				\caption{Box plot statistics (mean, standard deviation, maximum and minimum values) of the Euclidean position error of the prediction compared to the ground truth for different prediction algorithms and at particular prediction horizons}
				\label{fig:pred_error_vehicle1}
			\end{center}
		\end{figure}

\begin{figure}
			\begin{center}
				\includegraphics[scale=0.8]{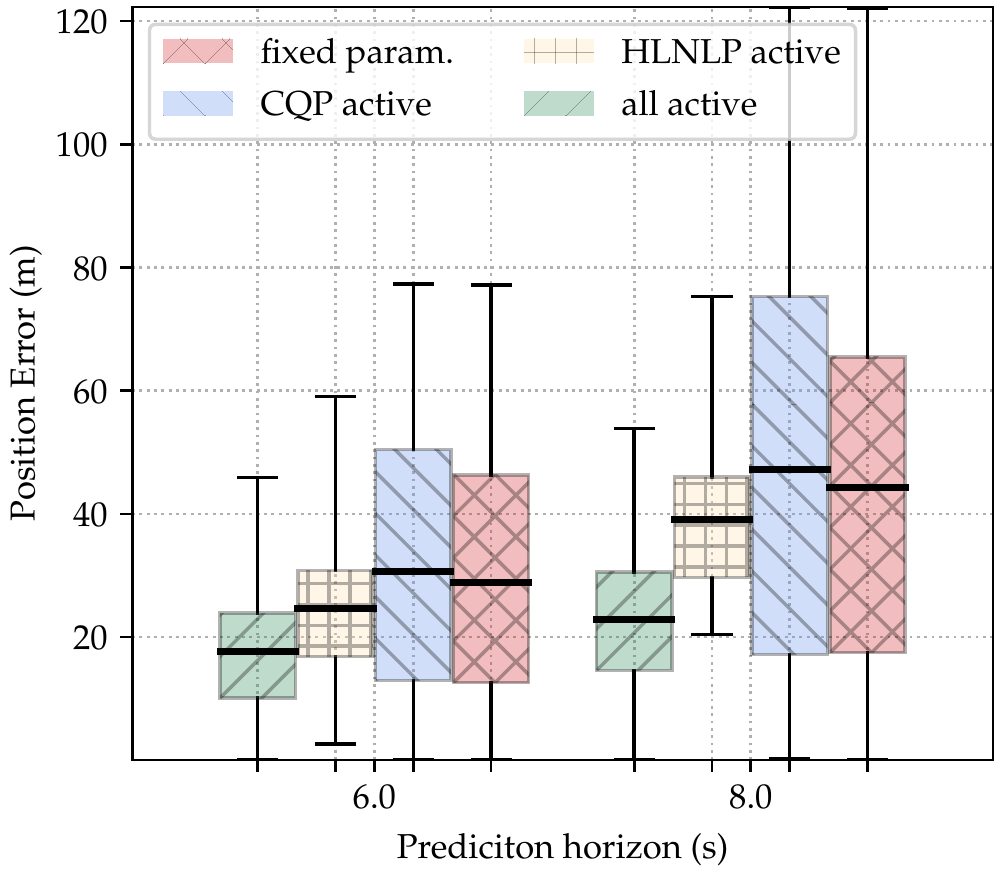}
				\caption{Box plot statistics (mean, standard deviation, maximum and minimum values) of the Euclidean position error of the prediction compared to the ground truth for different active components at a prediction horizon of 6 and 8 seconds.}
				\label{fig:full_estimated_error_modules}
			\end{center}
		\end{figure}

		The algorithm was evaluated with two opponent vehicles in a simulation environment as shown in Fig.~\ref{fig:racecar}. The two opponent race cars follow a racing line that was computed by a semi-analytic velocity profile computation as shown in \cite{Panagiotis2005} together with differently parameterized racing paths according to \cite{Reiter2021}. Therefore, the resulting trajectories are not in the solution space of the LLNLP and consequently can not be approximated exactly, which is similar to real observations. The HLNLP estimates the weight parameters and keeps converging to a semi-stationary solution after approximately 200 seconds as shown in Fig.~\ref{fig:full_estimated_weights}. The convergence behavior depends heavily on the choice of hyperparameters, particularly on the arrival weight $P$ in (\ref{eq:hlnlp_2}).
		After the weights converged, the predictions of all active components (\emph{all components}) were compared to other estimation algorithms. First, the initial parameter setting was simulated, where the weights and constraints were kept constant and only the LLNLP was active (\emph{LLNLP}). Secondly, a constant velocity estimation was used, where the path was computed by means of the PP component, but the velocity was set constant to the observed velocity (\emph{constant velocity}). Fig.~\ref{fig:pred_error_vehicle1} shows the comparison of the three settings by evaluating the Euclidean position error after certain prediction horizons. It can be seen that for increasing prediction horizons, the differences in the error measures becomes large, due to the acceleration constraints related to curves and the integrating errors. For short prediction horizons, the constant velocity estimator is performing similarly in our test cases, which was also observed in \cite{Schoeller2019}.
		The prediction performance with respect to the Euclidean position error was further compared by deactivating either the CQP or the HLNLP part. Fig.~\ref{fig:full_estimated_error_modules} shows the comparison with either components active (\emph{CQP active} or \emph{HLNLP active}), with all parameters fixed (\emph{fixed parameters}) or with the full algorithm (\emph{all active}) at a prediction horizon of 6 and 8 seconds.
		The results looked similar for all observed race cars. In our simulation the biggest improvement originated from the HLNLP part which can be seen in Fig.~\ref{fig:full_estimated_error_modules} and which is due to the rather aggressive driving behavior of the observed vehicles and the moderate initialization of the corresponding weight parameters of the LLNLP.

	\section{Conclusions}
		The paper presents a novel approach for predicting race car trajectories in real time and with sparse observation data. It is shown that the algorithm works in an embedded setting and yields satisfying predictions. The key advantage of using an optimization problem as a predictor is the natural integration of constraints. Nevertheless, the quality of the solution is restricted by the assumptions related to the low-level problem, e.g., which norms are used as penalties and what quantities are supposed to be penalized. The expressiveness of the low-level problem is limited due to its KKT conditions arising in a high-level optimization problem, which poses a non-smooth optimization problem with no guaranteed solution. Yet, in practice, the problem as stated is posed well enough to be solvable by means of a robust solver like \emph{IPOPT}. Future investigations might include an algorithm that also estimates an uncertainty measure and updates the arrival cost correspondingly and investigating rich function approximators in various parts of the algorithm to achieve a vanishing error, as the number of samples increases.
	
	\addtolength{\textheight}{-12cm}   

	\section*{ACKNOWLEDGMENT}
	This research was supported by DFG via Research Unit FOR 2401 and project 424107692 and by the EU via ELO-X 953348.
	The authors thank Jonathan Frey, Katrin Baumgärtner, Jasper Hoffmann, Martin Kirchengast and the other members of the \emph{Autonomous Racing Graz} team for their valuable input to this work.
	
	\bibliographystyle{IEEEtran} 
	\bibliography{IEEEabrv,library,syscop}

\begin{thebibliography}{10}
\providecommand{\url}[1]{#1}
\csname url@samestyle\endcsname
\providecommand{\newblock}{\relax}
\providecommand{\bibinfo}[2]{#2}
\providecommand{\BIBentrySTDinterwordspacing}{\spaceskip=0pt\relax}
\providecommand{\BIBentryALTinterwordstretchfactor}{4}
\providecommand{\BIBentryALTinterwordspacing}{\spaceskip=\fontdimen2\font plus
\BIBentryALTinterwordstretchfactor\fontdimen3\font minus
  \fontdimen4\font\relax}
\providecommand{\BIBforeignlanguage}[2]{{%
\expandafter\ifx\csname l@#1\endcsname\relax
\typeout{** WARNING: IEEEtran.bst: No hyphenation pattern has been}%
\typeout{** loaded for the language `#1'. Using the pattern for}%
\typeout{** the default language instead.}%
\else
\language=\csname l@#1\endcsname
\fi
#2}}
\providecommand{\BIBdecl}{\relax}
\BIBdecl

\bibitem{Florin2021}
\BIBentryALTinterwordspacing
F.~Leon and M.~Gavrilescu, ``A review of tracking and trajectory prediction
  methods for autonomous driving,'' \emph{Mathematics}, vol.~9, no.~6, 2021.
  [Online]. Available: \url{https://www.mdpi.com/2227-7390/9/6/660}
\BIBentrySTDinterwordspacing

\bibitem{Millefiori2021}
\BIBentryALTinterwordspacing
S.~Capobianco, L.~M. Millefiori, N.~Forti, P.~Braca, and P.~Willett, ``Deep
  learning methods for vessel trajectory prediction based on recurrent neural
  networks,'' \emph{CoRR}, vol. abs/2101.02486, 2021. [Online]. Available:
  \url{https://arxiv.org/abs/2101.02486}
\BIBentrySTDinterwordspacing

\bibitem{ZHANG20209}
\BIBentryALTinterwordspacing
J.~Zhang, H.~Liu, Q.~Chang, L.~Wang, and R.~X. Gao, ``Recurrent neural network
  for motion trajectory prediction in human-robot collaborative assembly,''
  \emph{CIRP Annals}, vol.~69, no.~1, pp. 9--12, 2020. [Online]. Available:
  \url{https://www.sciencedirect.com/science/article/pii/S0007850620300998}
\BIBentrySTDinterwordspacing

\bibitem{Andre2021}
A.~Ip, L.~Irio, and R.~Oliveira, ``Vehicle trajectory prediction based on lstm
  recurrent neural networks,'' in \emph{2021 IEEE 93rd Vehicular Technology
  Conference (VTC2021-Spring)}, 2021, pp. 1--5.

\bibitem{nikhil2018convolutional}
N.~Nikhil and B.~T. Morris, ``Convolutional neural network for trajectory
  prediction,'' 2018.

\bibitem{zanon2020}
\BIBentryALTinterwordspacing
S.~Gros and M.~Zanon, ``Data-driven economic nmpc using reinforcement
  learning,'' \emph{IEEE Transactions on Automatic Control}, vol.~65, no.~2, p.
  636–648, Feb 2020. [Online]. Available:
  \url{http://dx.doi.org/10.1109/TAC.2019.2913768}
\BIBentrySTDinterwordspacing

\bibitem{Kolter2019}
\BIBentryALTinterwordspacing
A.~Agrawal, B.~Amos, S.~T. Barratt, S.~P. Boyd, S.~Diamond, and J.~Z. Kolter,
  ``Differentiable convex optimization layers,'' \emph{CoRR}, vol.
  abs/1910.12430, 2019. [Online]. Available:
  \url{http://arxiv.org/abs/1910.12430}
\BIBentrySTDinterwordspacing

\bibitem{lucidgames}
\BIBentryALTinterwordspacing
S.~L. Cleac'h, M.~Schwager, and Z.~Manchester, ``Lucidgames: Online unscented
  inverse dynamic games for adaptive trajectory prediction and planning,''
  \emph{CoRR}, vol. abs/2011.08152, 2020. [Online]. Available:
  \url{https://arxiv.org/abs/2011.08152}
\BIBentrySTDinterwordspacing

\bibitem{algames}
\BIBentryALTinterwordspacing
S.~Le~Cleac’h, M.~Schwager, and Z.~Manchester, ``Algames: A fast solver for
  constrained dynamic games,'' \emph{Robotics: Science and Systems XVI}, Jul
  2020. [Online]. Available: \url{http://dx.doi.org/10.15607/RSS.2020.XVI.091}
\BIBentrySTDinterwordspacing

\bibitem{Menner2021}
M.~Menner, P.~Worsnop, and M.~N. Zeilinger, ``Constrained inverse optimal
  control with application to a human manipulation task,'' \emph{IEEE
  Transactions on Control Systems Technology}, vol.~29, no.~2, pp. 826--834,
  2021.

\bibitem{Mombaur2010}
K.~Mombaur, A.~Truong, and J.-P. Laumond, ``From human to humanoid
  locomotion-an inverse optimal control approach,'' \emph{Auton. Robots},
  vol.~28, pp. 369--383, 04 2010.

\bibitem{Menner2018ConvexFA}
M.~Menner and M.~N. Zeilinger, ``Convex formulations and algebraic solutions
  for linear quadratic inverse optimal control problems,'' \emph{2018 European
  Control Conference (ECC)}, pp. 2107--2112, 2018.

\bibitem{Sinha2018}
A.~Sinha, P.~Malo, and K.~Deb, ``A review on bilevel optimization: From
  classical to evolutionary approaches and applications,'' \emph{IEEE
  Transactions on Evolutionary Computation}, vol.~22, no.~2, pp. 276--295,
  2018.

\bibitem{Reiter2021}
R.~Reiter and M.~Diehl, ``Parameterization approach of the frenet
  transformation for model predictive control of autonomous vehicles,''
  \emph{Proceedings of the European Control Conference (ECC)}, 2021.

\bibitem{Reiter2021a}
R.~Reiter, M.~Kirchengast, D.~Watzenig, and M.~Diehl, ``Mixed-integer
  optimization-based planning for autonomous racing with obstacles and
  rewards,'' \emph{Proceedings of the IFAC Conference on Nonlinear Model
  Predictive Control (NMPC)}, 2021.

\bibitem{Verschueren2019}
\BIBentryALTinterwordspacing
R.~Verschueren, G.~Frison, D.~Kouzoupis, J.~Frey, N.~van Duijkeren, A.~Zanelli,
  B.~Novoselnik, J.~Frey, T.~Albin, R.~Quirynen, and M.~Diehl, ``acados: a
  modular open-source framework for fast embedded optimal control,''
  \emph{arXiv preprint; accepted at: Mathematical Programming Computation},
  2019. [Online]. Available: \url{https://arxiv.org/abs/1910.13753}
\BIBentrySTDinterwordspacing

\bibitem{Rawlings2017}
J.~B. Rawlings, D.~Q. Mayne, and M.~M. Diehl, \emph{Model Predictive Control:
  Theory, Computation, and Design}, 2nd~ed.\hskip 1em plus 0.5em minus
  0.4em\relax Nob Hill, 2017.

\bibitem{Waechter2006}
A.~W\"achter and L.~T. Biegler, ``On the implementation of an interior-point
  filter line-search algorithm for large-scale nonlinear programming,''
  \emph{Mathematical Programming}, vol. 106, no.~1, pp. 25--57, 2006.

\bibitem{Andersson2019}
J.~A.~E. Andersson, J.~Gillis, G.~Horn, J.~B. Rawlings, and M.~Diehl,
  ``{CasADi} -- a software framework for nonlinear optimization and optimal
  control,'' \emph{Mathematical Programming Computation}, vol.~11, no.~1, pp.
  1--36, 2019.

\bibitem{osqp}
\BIBentryALTinterwordspacing
B.~Stellato, G.~Banjac, P.~Goulart, A.~Bemporad, and S.~Boyd, ``{OSQP}: an
  operator splitting solver for quadratic programs,'' \emph{Mathematical
  Programming Computation}, vol.~12, no.~4, pp. 637--672, 2020. [Online].
  Available: \url{https://doi.org/10.1007/s12532-020-00179-2}
\BIBentrySTDinterwordspacing

\bibitem{Schoeller2019}
\BIBentryALTinterwordspacing
C.~Sch{\"{o}}ller, V.~Aravantinos, F.~Lay, and A.~C. Knoll, ``The simpler the
  better: Constant velocity for pedestrian motion prediction,'' \emph{CoRR},
  vol. abs/1903.07933, 2019. [Online]. Available:
  \url{http://arxiv.org/abs/1903.07933}
\BIBentrySTDinterwordspacing

\bibitem{Panagiotis2005}
E.~Velenis and P.~Tsiotras, ``Optimal velocity profile generation for given
  acceleration limits: Theoretical analysis,'' 07 2005, pp. 1478 -- 1483 vol.
  2.

\end{thebibliography}
\end{document}